\documentclass[conference,letterpaper]{IEEEtran}
\def\IEEEtitletopspace{24pt}

\IEEEoverridecommandlockouts
\usepackage{cite}
\usepackage{amsmath,amssymb,amsfonts}
\usepackage{algorithmic}
\usepackage{graphicx}
\usepackage{textcomp}
\usepackage{xcolor}
\usepackage{cite}
\usepackage{amsmath,amssymb,amsfonts}
\usepackage{graphicx}
\usepackage{textcomp}
\usepackage{xcolor}
\usepackage{algorithm}
\usepackage{algorithmic}
\usepackage{tikz}
\usepackage{enumitem}
\usepackage{inconsolata}
\usepackage{threeparttable}
\usepackage{multirow}
\usepackage{comment}
\usepackage{csquotes}
\usepackage{amsthm}
\usepackage{subcaption}

\newtheoremstyle{italiclabel} 
  {\topsep}   
  {\topsep}   
  {\normalfont}  
  {}          
  {\itshape}  
  {:}         
  {.5em}      
  {}          

\theoremstyle{italiclabel}

\newtheorem{remark}{Remark}

\usepackage[font=normal]{caption} 
\usepackage{fancyhdr}
\usepackage{caption}
\usepackage{tikz}
\usetikzlibrary{shapes,arrows,positioning,calc}
\tikzset{
block/.style = {draw, fill=white, rectangle, minimum height=2.5em, minimum width=3em},
tmp/.style = {coordinate},
sum/.style= {draw, fill=white, circle, node distance=1cm},
input/.style = {coordinate},
output/.style= {coordinate},
pinstyle/.style = {pin edge={to-,thin,black}}}

\usepackage{svg}
\usepackage[off]{svg-extract}
\svgsetup{clean=true}

\def\BibTeX{{\rm B\kern-.05em{\sc i\kern-.025em b}\kern-.08em
    T\kern-.1667em\lower.7ex\hbox{E}\kern-.125emX}}
\begin{document}

\IEEEsettopmargin{t}{54pt}
\IEEEsettextwidth{56pt}{56pt}
\IEEEsetsidemargin{c}{0pt}

\title{A Neural Network-based Multi-timestep Command Governor for Nonlinear Systems with Constraints \\
}

\author{
    Mostafaali Ayubirad, \textit{Member, IEEE}, and Hamid R. Ossareh, \textit{Senior Member, IEEE}%
    \thanks{This material is based upon work supported by the U.S. Department of Energy (DOE) under award number DE-EE0010147, and Ford Motor Company under award 002092-URP. The views expressed herein do not necessarily represent the views of the DOE, the United States Government, or Ford Motor Company.}%
    \thanks{The authors are with the Department of Electrical and Biomedical Engineering, University of Vermont, Burlington, VT 05405 USA (e-mail: {\tt mostafa-ali.ayubirad@uvm.edu;
hamid.ossareh@uvm.edu)}.}
}

\maketitle

\begin{abstract}

The multi-timestep command governor (MCG) is an add-on algorithm that enforces constraints by modifying, at each timestep, the reference command to a pre-stabilized control system. The MCG can be interpreted as a Model-Predictive Control scheme operating on the reference command.
The implementation of MCG on nonlinear systems carries a heavy computational burden as it requires solving a nonlinear program with multiple decision variables at each timestep. This paper proposes a less computationally demanding alternative, based on approximating the MCG control law using a neural network (NN) trained on offline data. However, since the NN output may not always be constraint-admissible due to training errors, its output is adjusted using a sensitivity-based method.  
We thus refer to the resulting control strategy as the neural network-based MCG (NN-MCG). As validation, the proposed controller is applied as a load governor for
constraint management in an automotive fuel cell system. It is shown that the proposed strategy is significantly more computationally efficient than the traditional MCG, while achieving nearly identical performance if the NN is well-trained.
\end{abstract}


\section{Introduction}
Constraint management of nonlinear control systems  has received significant attention over the past few decades. Model predictive control (MPC) is one of the advanced control strategies that has been widely adopted in industry for controlling nonlinear processes with system constraints \cite{qin2003survey}. However, practical application of MPC in fast nonlinear dynamical systems is challenging due to the high computational cost of its online optimization, see for example \cite{shi2017advanced}. 

As alternatives to MPC, Reference Governors (RG) and Command Governors (CG) \cite{gilbert1995discrete,bemporad1997nonlinear} offer more computationally efficient solutions, though at the cost of some performance loss. Governor schemes serve as add-on mechanisms for closed-loop systems, adjusting the reference command in real time to enforce state and input constraints when necessary. Another appealing feature of governor schemes for practitioners is the decoupling of tracking/stabilization from constraint management. Some application of RG can be found in automotive fuel cell \cite{ayubirad2023simultaneous}, and aerospace systems \cite{frey2016time}. 

A common feature of RG and CG schemes is that, at each time instant, they compute a modified reference command that, if held constant from that point forward, satisfies the constraints for all time. 
The difference between them is that CG solves a quadratic program, but RG further constrains the input space and thus solves a linear program, which is more computationally efficient. A generalization of RG and CG is the so-called Extended Command Governor (ECG), which improves performance over RG and CG by employing a time-varying command with prescribed dynamics instead of a constant one \cite{kalabic2012reduced}. Another extension was proposed in \cite{vahidi2005constraint}, where the governor scheme was formulated as an MPC  applied to the closed-loop system without assuming the prescribed input dynamics of ECG. 
In this paper, we refer to the scheme in \cite{vahidi2005constraint} as the multi-timestep command governor (MCG). While the MCG offers a larger decision space than the other approaches, its optimization problem is more complex.

All of the above approaches have been developed for linear closed-loop systems, but can be extended to nonlinear systems by incorporating the nonlinear dynamics into the optimization problem. However, this results in a nonlinear program (NLP), which, like nonlinear MPC (NMPC) \cite{allgower2012nonlinear}, is computationally demanding and often infeasible for real-time implementation.  The literature has addressed this issue through constraint tightening,  Lyapunov methods, and/or constraining the input space, see e.g., \cite{bemporad1998reference,vahidi2006constraint,gilbert2002nonlinear,osorio2022novel}. In this paper, we provide an alternative approach to further enhance the computational footprint of MCG without constraining the input space or using heuristics.



\begin{figure}
    \centering
   \begin{subfigure}{0.4\textwidth}
        \centering
        \includegraphics[width=1.0\textwidth]{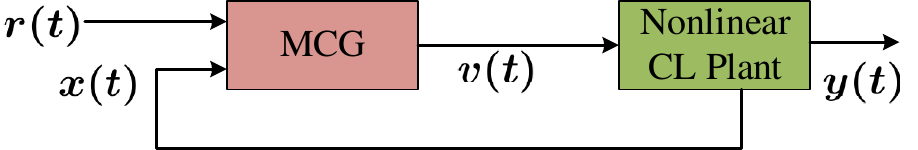}
        \caption{MCG}
    \end{subfigure}
    \begin{subfigure}{0.5\textwidth}
        \vspace{0.5em}
\includegraphics[width=1.0\textwidth,height=0.20\textwidth]{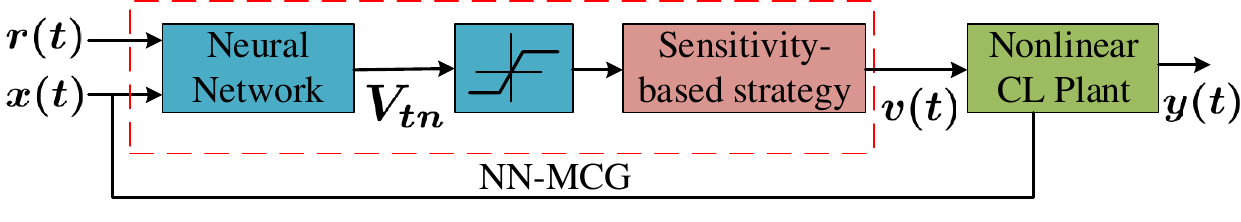}
        \caption{NN-MCG}
    \end{subfigure} 
    \caption{constraint management schemes. The signals are as follows: $y(t)$ is the constrained output, $r(t)$ is the desired reference, $v(t)$ is the modified reference command, $V_{tn}(t)$ is the nominal input sequence, and $x(t)$ is the system state (measured or estimated).}
    \label{fig:RG schemes}
    \vspace{-1.2em}
\end{figure}

Our approach is based on approximating the MCG control law using a regression neural network (NN) trained on offline data collected from a closed-loop system governed by the MCG (see Fig.\ref{fig:RG schemes}(a)). Since approximation errors can lead to constraint violations, several approaches have been proposed to address this issue in the context of NMPC. These include probabilistic methods without deterministic guarantees of constraint satisfaction~\cite{hertneck2018learning}, safety filters that rely on control-invariant set computations that are intractable for nonlinear systems~\cite{wabersich2021predictive}, and fallback inputs that trade off optimality for simplicity~\cite{hose2023approximate}. In this work, we adjust the NN output using a sensitivity-based method to enhance the constraint
enforcement capability of the NN, enabling both real-time feasibility and near-optimal performance (see Fig.\ref{fig:RG schemes}(b)).
We refer to the proposed strategy as neural network-based MCG (NN-MCG). To evaluate its effectiveness and compare its performance with MCG in both constraint enforcement and tracking performance, we apply NN-MCG as a load governor for an automotive fuel cell (FC) model described in \cite{pukrushpan2004control}. Specifically, we use NN-MCG to enforce compressor surge, compressor choke, and oxygen starvation. We show that with a well-trained NN, the NN-MCG performs almost identically to MCG, but is 23 times faster in the worst-case scenario.

In summary, the main contributions of this article are:

\begin{itemize}
    \item The formulation of the NN-MCG control strategy for constraint management of nonlinear systems;
    \item The development of a computationally efficient load governor based on NN-MCG to enforce the constraints in a FC system;
    \item Comparison of the computational footprint and performance of NN-MCG against MCG and NN solutions.
\end{itemize}

The rest of this article is organized as follows. Section~\ref{sec:SYSTEM DESCRIPTION} contains the system description and assumptions. Section~\ref{sec:MPC-RG introduction} introduces the MCG. The novel NN-MCG is proposed in Section~\ref{sec:NN-MCG}. Section~\ref{sec:Implementation of the modified NNRG on the FC air-path system} presents the NN-MCG application to FC system. Finally, conclusions are drawn in Section~\ref{sec:conclusions}.

\section{System description and Assumptions}
\label{sec:SYSTEM DESCRIPTION}
Consider Fig.~\ref{fig:RG schemes}, in which the ``Nonlinear CL Plant'' represents a general closed-loop nonlinear system whose dynamics are described by
\begin{flalign}
        x(t+1)=f(x(t),v(t))
\label{eqn:NL_dynamic}
\end{flalign}
where $x(t)\in {\mathbb{R}}^{n_x}$ denotes the state vector, incorporating both the plant and controller states; $v(t)\in \mathbb{R}$ is the modified reference command, and the constrained output $y(t)\in {\mathbb{R}}$ is subjected to the following  constraint:
\begin{flalign}
y(t)=h(x(t),v(t))\le 0
\label{eqn:NL_constraint}
\end{flalign}
We assume that the system is bounded-input bounded-state stable. We also assume that the input satisfies \( v(t) \in \mathcal{V} \), where \( \mathcal{V} \) is compact. This, together with the stability assumption, implies that the state is also bounded, i.e., \( x(t) \in \mathcal{X} \) for some compact set \( \mathcal{X} \). Moreover, the functions \( f(x, v) \) and \( h(x, v) \) are assumed to be twice continuously-differentiable over the domain \( \mathcal{X} \times \mathcal{V} \). This condition ensures the existence of sensitivity functions and guarantees a bounded remainder term in the first-order Taylor series expansion discussed in Section~\ref{sec:NN-MCG}. Finally, we assume that \( \forall x(0) \in \mathcal{X} \) and constant input \( v \in \mathcal{V} \), the equilibrium manifolds \( \bar{x}_v = \lim_{t \to \infty} x(t) \) and \( \bar{y}_v = \lim_{t \to \infty} y(t) \), which depend only on $v$, exist and are uniquely defined. 

\section{Overview of the Multi-timestep command governor (MCG)}
\label{sec:MPC-RG introduction}

This section provides an overview of the MCG (see \cite{vahidi2005constraint}) for nonlinear systems. 

Consider the block diagram in Fig.~\ref{fig:RG schemes}(a). At each timestep \( t \), the MCG 
computes an optimal control sequence ${V_t = [v_t(0), v_t(1), \dots, v_t(N)]^\top}$ over a prediction horizon $N$, where \( v_t(j) \) denotes the modified reference at time \( t+j \) computed at the current time \( t \). This sequence is chosen to minimize its deviation from 
$r(t)$ while enforcing the constraint in \eqref{eqn:NL_constraint} over the prediction horizon.


More specifically, let \( x(t+j|t) \) and \( y(t+j|t) \), or in short \( x_t(j) \) and \( y_t(j) \), respectively, denote the state and output at time \( t+j \) predicted at time \( t \). The MCG computes the optimal input sequence by solving the following finite-horizon discrete-time optimal control problem:
\begin{flalign}
    \begin{aligned}
        \min_{\varepsilon,V_t} \quad & \sum_{j=0}^N \left(r(t) - v_t(j)\right)^2 + \rho_s \sum_{j=0}^{N-1} \left(v_t(j) - v_t(j+1)\right)^2 \\
        & + \rho \varepsilon^2 \\
        \text{s.t.} \quad & y_t(j) = h(x_t(j), v_t(j)) \leq \varepsilon, \quad j = 0, \ldots, N, \\
        & x_t(j+1) = f(x_t(j), v_t(j)), \quad j = 0, \ldots, N-1, \\
        & x_t(0) = x(t), \quad \varepsilon \geq 0,
    \end{aligned}
\label{eqn:MCG optimization}
\end{flalign}
where we have assumed that a full measurement or an accurate estimate of the state \( x(t) \) is available at the current time \( t \). 
The cost function in \eqref{eqn:MCG optimization} consists of three terms: 1) the deviation of the modified reference \( v_t(j) \) from the desired setpoint \( r(t) \); 2) a penalty term on the rate of change of the modified reference \( v_t(j) \), weighted by \( \rho_s \), which smooths transitions between consecutive control inputs; and 3) a penalty on the slack variable \( \varepsilon \), weighted by \( \rho \). The slack variable relaxes (i.e., softens) the output constraint in \eqref{eqn:NL_constraint} to eliminate feasibility issues and prevent governor ``hang-up'', wherein the governor stops updating the command.  We choose $\rho$ to be large to strongly penalize constraint violations. 

\begin{remark}
    Note that if $\varepsilon=0$ is feasible, then \eqref{eqn:MCG optimization} yields a solution with $\varepsilon=0$ at optimality, ensuring constraint satisfaction over the prediction horizon. However, this does not necessarily guarantee constraint satisfaction over an infinite horizon, even under a perfect model. Infinite-horizon constraint enforcement can be achieved by imposing additional constraints on the terminal state and/or selecting a sufficiently large horizon length 
$N$, as is common in the  MPC literature. However, due to inevitable plant-model mismatches in practice, the inclusion of \( \varepsilon \) is still recommended.
\end{remark}

\begin{remark} 
Note that the optimization problem in \eqref{eqn:MCG optimization} is just one way to extend MCG to nonlinear dynamics. Alternative formulations can also be considered, such as those ensuring infinite-horizon constraint enforcement using terminal sets. The approach presented in Section~\ref{sec:NN-MCG} can be extended to these formulations as well.
\end{remark}

Let $V^*_t = [v_t^*(0), \dots, v_t^*(N)]^\top$ be the optimal solution of \eqref{eqn:MCG optimization} at time \( t \). Similar to MPC, we apply its first element to the system:
\vspace{-0.5em}
\begin{flalign}
v(t)=v^*_t(0)
\label{eqn:v_t optimal}
\end{flalign}
The optimization problem \eqref{eqn:MCG optimization} is solved again at time \( t+1 \), using the updated state \( x(t+1) \), resulting in a receding horizon control approach.


The MCG can be seen as a NMPC with two key differences: (\textit{i}) while NMPC computes control inputs directly within the feedback loop, the MCG modifies the reference \( r \) before it enters the loop; and (\textit{ii}) MCG penalizes \( r - v \) instead of the NMPC's objective, which penalizes the output or state tracking error and the control effort. The rest of the formulation  
(e.g., penalty on control-input increments and the use of nonlinear dynamics) is similar to standard NMPC.
As a result, The MCG retains properties similar to  NMPC, which we will not delve into further.

The MCG algorithm requires solving an online NLP problem at each timestep, which is computationally expensive. To enhance computational efficiency, the NLP can be warm-started using the shifted optimal solution from the previous timestep, as is standard practice in MPC. 
Although warm start can help the solver converge to an optimal solution faster, the computational cost of MCG still poses significant challenges for real-time implementation on fast dynamic systems. In the next section, we adopt a neural network approximation followed by a novel sensitivity-based approach to impose the soft constraint in \eqref{eqn:MCG optimization} while ensuring computational viability, even in resource-constrained environments.

\section{Neural network-based multi-timestep command governor (NN-MCG)}
\label{sec:NN-MCG}
This section first introduces, in Subsection IV-A, a regression NN, which is trained offline to approximate the input-output map of the MCG. This allows us to approximate the optimal input sequence $V_t^*$ at every timestep in a computationally efficient manner. In subsection IV-B, we use this input sequence to predict the output trajectory of the nonlinear system over the prediction horizon and then adjust the NN output using a sensitivity-based method to enforce the constraints. We refer to the entire scheme as NN-MCG (see Fig.~\ref{fig:RG schemes} (b)).


\subsection{Neural network: Data collection and training}
NNs are universal function approximators capable of approximating any continuous function with reasonable accuracy. 
Thus, the idea here is to use an NN to approximate the optimal control sequence $V_t^*$ of the MCG. 

Let $\psi([x(t); r(t)])$ represent the implicit function of MCG that maps the current state and reference, \([x(t); r(t)]\), to the optimal input sequence, $V_t^*$. This mapping is typically quite complex, so we adopt a feedforward NN to approximate it. Let $\psi_{NN}([x(t); r(t)])$ denote the input-output map of a feedforward NN with $L$ layers, that is:
\begin{flalign}
\psi_{{NN}}(\cdot) = g^{(L)}\left(g^{(L-1)}\left(\dots g^{(2)}\left(g^{(1)}(\cdot)\right)\right)\right)
\label{eqn:NN_function}
\end{flalign}
where $g^{(l)} = \sigma^{(l)}\left(W^{(l)} g^{(l-1)} + b^{(l)}\right), l = 1, \dots, L$. Here, $\sigma^{(l)}(\cdot)$ is the activation function of the $l$-th layer in NN, $W^{(l)}$ and $b^{(l)}$ are the weights and biases of each layer, and $g^{(0)}$  corresponds to \( [x(t); r(t)] \). Note that ideally, $\psi_{NN}(\cdot)$ would equal $\psi(\cdot)$ for all $x(t)$ and $r(t)$; however, due to inherent approximation errors of NNs, this is not always the case.

To train the NN (i.e., determine the best weights and biases), we collect a  dataset of inputs \( [x(t); r(t)] \) and their corresponding outputs $V_t^*$, generated by executing the MCG on the nonlinear system across diverse scenarios (i.e., various reference profiles, $r(t)$, as observed in practice). This ensures that the supervised trained NN achieves strong generalization performance, leading to a  reasonable approximation of $V_t^*$:
\begin{flalign}
V_{tn}=\psi_{NN}([x(t); r(t)]) 
\label{eqn:v_tn optimal}
\end{flalign}
where $V_{tn} = [v_{tn}(0), \dots, v_{tn}(N)]^\top$ is the approximated control input through the NN at time $t$. 

During real-time operations, the trained NN is evaluated at each timestep to obtain \( V_{tn} \). The elements of \( V_{tn} \) are then saturated to ensure \( v_{tn}(j) \in \mathcal{V} \) and passed to a sensitivity-based strategy, which will be discussed in the next section. This process is represented by NN and the saturation block in Fig.~\ref{fig:RG schemes}(b).

In summary, data collection and NN training in NN-MCG consist of the following steps:
\begin{itemize}
    \item \textbf{Offline data collection}: Execute the MCG on the nonlinear system with a given reference signal \( r(t) \) that covers a broad range of operating conditions of the system, either through hardware experiments or system simulations if a simulator is available, to create a training dataset for the NN.
    \item \textbf{Offline training}: Train multiple NNs with different architectures, varying the number of hidden layers and neurons per layer. Conduct multiple trials for each architecture, randomizing initial weights and biases in \eqref{eqn:NN_function}. Following the standard approaches \cite{geron2022hands}, select the architecture achieving the best performance, evaluated using metrics like root-mean square error (RMSE), to acquire the NN control law in \eqref{eqn:v_tn optimal}.
    \item \textbf{Online operation}: Use the NN control law to approximate the optimal input sequence, saturate its elements as discussed above, and pass them to the sensitivity-based strategy as discussed in the next section.
\end{itemize}

\remark{Due to the inherent approximation errors of the NN, applying the first element of $V_{tn}$ cannot guarantee constraint satisfaction. We refer to this approach as ``naive NN" and demonstrate its failure compared to our proposed approach, NN-MCG, in Section~\ref{sec:Implementation of the modified NNRG on the FC air-path system}. This is the reason why we employ a sensitivity-based method to further enhance the constraint enforcement capability of the NN.
\label{re:NN_fail}
  }

\subsection{Sensitivity-based strategy for online adjustment of neural network output}
\label{sec:sensitivity}


The high-level idea of the sensitivity-based strategy in NN-MCG involves using the saturated NN output to solve Eqs. \eqref{eqn:NL_dynamic} and  \eqref{eqn:NL_constraint} and thus obtain a ``nominal predicted output trajectory'' over the prediction horizon.  Using this nominal trajectory, we compute a truncated Taylor series expansion of the output prediction around the nominal input sequence $V_{tn}$. The truncation error is then bounded, and this bound is used to tighten the constraint in \eqref{eqn:MCG optimization} to make it robust against the worst-case truncation errors. This results in an optimization problem similar to \eqref{eqn:MCG optimization}, but it takes the form of a quadratically constrained quadratic program (QCQP), which is easier to solve than a general nonlinear program. 
Below, we elaborate on these ideas and then discuss how the remainder term can be bounded in practice.

To begin, note that, at each time instant \( t \), the nominal output, starting from the initial condition \( x(t) \), is computed by solving the following difference equation:
\begin{flalign}
    \begin{split}
        {x}_{tn}(j+1) &= f({x}_{tn}(j), v_{tn}(j)) \\ 
        {y}_{tn}(j) &= h({x}_{tn}(j), v_{tn}(j)) \\
        {x}_{tn}(0) &= x(t)
    \end{split}
\label{eqn:nominal trajectory}
\end{flalign}
where $j = 0, \ldots, N$. Since ${y}_t(j)$ depends on the values in $V_{tn}$, we expand ${y}_t(j)$ in a Taylor series around the nominal input sequence $V_{tn}$:
\begin{align}
    y_t(j) = y_{tn}(j) + \sum_{k=0}^{j} S_y(j, k) \left(v_t(k) - v_{tn}(k)\right) + R_t(j)
    \label{eqn:taylor expansion}
\end{align}
where $S_y(j, k) = \frac{dy_t(j)}{dv_t(k)}$ is the sensitivity of ${y_t(j)}$ with respect to $v_t(k)$ ($k \le j$), evaluated at $v_{tn}(j)$ along the nominal predicted state trajectories ${x}_{tn}(j)$, and the term $R_t(j)$ is the remainder term in the Taylor series. 
According to the law of total derivatives, the $S_{y}(j,k)$ in (\ref{eqn:taylor expansion}) is calculated as:
\begin{flalign}
\begin{split}
S_y(j, k) = & \frac{\partial h(x_t(j), v_t(j))}{\partial x_t(j)} S_x(j, k) + \frac{\partial h(x_t(j), v_t(j))}{\partial v_t(k)}
\end{split}
\label{eqn:y_sensitivity}
\end{flalign}
where the partial derivatives are evaluated at $x_{tn}(j)$ and $v_{tn}(j)$. Note that the partial derivative of $h(x_t(j), v_t(j))$ with respect to $v_t(k)$ is $1$ when $j=k$, and $0$ otherwise. $S_x(j,k)$  is the sensitivity of $x_t(j)$ with respect to $v_t(k)$, which can be computed as a solution to the following difference equation:
\begin{flalign}
\begin{split}
S_x(j+1, k) = & \frac{\partial f(x_t(j), v_t(j))}{\partial x_t(j)} 
 S_x(j, k)  + \frac{\partial f(x_t(j), v_t(j))}{\partial v_t(k)} 
 \\&
        S_x(0, k) = 0, \quad \forall k.
\end{split}
\label{eqn:x_sensitivity}
\end{flalign}
where the partial derivatives are evaluated at $x_{tn}(j)$ and $v_{tn}(j)$. Recall that, as stated in Section~\ref{sec:SYSTEM DESCRIPTION}, the functions \( f \) and \( h \) are twice continuously differentiable on the compact set \( \mathcal{X} \times \mathcal{V} \). As a result, \( \frac{d^2 y}{dv^2} \) is continuous on this domain and satisfies the conditions of the Extreme Value Theorem. Therefore, there exists a constant \( M \geq 0 \) such that \( \sup_{(x, v) \in \mathcal{X} \times \mathcal{V}} \left| \frac{d^2 y}{dv^2} \right| \leq M \). By Taylor's theorem, the cumulative remainder term in \eqref{eqn:taylor expansion} satisfies $${\left|R_t(j)\right| \le \frac{M}{2} \sum_{k=0}^{j} \left(v_t(k) - v_{tn}(k)\right)^2}.$$ For now, we assume that $M$ is known. The  computation of $M$ in the context of NN-MCG will be discussed later. With this upper bound on \( R_t(j) \), the output over the prediction horizon can be upper bounded as: 
\vspace{-0.2em}
\begin{flalign}
    \begin{split}
        {y}_t(j) \leq & \; y_{tn}(j) + \sum_{k=0}^{j} S_y(j, k) \left(v_t(k) - v_{tn}(k)\right) \\
        & + \frac{M}{2} \sum_{k=0}^{j} \left(v_t(k) - v_{tn}(k)\right)^2, \quad j = 0, \ldots, N
    \end{split}
\label{eqn:y_upper_prediction}
\end{flalign}
Thus, the soft constraint \({y}_t(j) \leq \varepsilon\) over the prediction horizon can be replaced by the following tightened condition:
\begin{flalign}
   \begin{split}
       & y_{tn}(j) + \sum_{k=0}^{j} S_y(j, k)  \left(v_t(k) - v_{tn}(k)\right)\\
       & + \frac{M}{2} \sum_{k=0}^{j} \left(v_t(k) - v_{tn}(k)\right)^2\le \varepsilon, \quad j = 0, \ldots, N
   \end{split}
\label{eqn:y upper bound replacment}
\end{flalign}
Because of \eqref{eqn:y_upper_prediction}, if $V_t$ satisfies \eqref{eqn:y upper bound replacment}, then ${y_t}(j)\leq \varepsilon$  and constraint are enforced as soft constraints over the prediction horizon. 
Thus, the ``Sensitivity-based strategy'' in Fig.~\ref{fig:RG schemes}(b) is formulated as the following optimization problem, which is a modification of \eqref{eqn:MCG optimization} using the above approximation:
\begin{flalign}
    \begin{aligned}
        \min_{\varepsilon, V_t} \quad & \sum_{j=0}^N \left(r(t) - v_t(j)\right)^2 + \rho_s \sum_{j=0}^{N-1} \left(v_t(j) - v_t(j+1)\right)^2 \\
        & + \rho \varepsilon^2 \\ 
        \text{s.t.} \quad & y_{tn}(j) + \sum_{k=0}^{j} S_y(j, k)  \left(v_t(k) - v_{tn}(k)\right)\\
       & + \frac{M}{2} \sum_{k=0}^{j} \left(v_t(k) - v_{tn}(k)\right)^2 \leq \varepsilon, \quad j = 0, \ldots, N, \\
        & \varepsilon \geq 0.
    \end{aligned}
\label{eqn:NN-MCG optimization}
\end{flalign}
Upon finding the optimal control input $V_t$, the first element is applied as the one-step reference for the closed-loop system. Note that the optimization problem in \eqref{eqn:NN-MCG optimization} reduces to a quadratic programming (QP) problem when \( M = 0 \) and a convex QCQP when \( M \neq 0 \), both of which are computationally more efficient than the NLP in \eqref{eqn:MCG optimization}.
\begin{remark}
    We remark on the connection between \eqref{eqn:NN-MCG optimization} and
\eqref{eqn:MCG optimization}. Since \eqref{eqn:NN-MCG optimization} uses the upper bound of output prediction in its optimization, its feasible set is strictly smaller than that of \eqref{eqn:MCG optimization}, which means that \eqref{eqn:NN-MCG optimization} yields a suboptimal solution of \eqref{eqn:MCG optimization}. If the NN is well-trained, then the feasible sets are almost the same and therefore the optimization problems yield similar objectives and thus similar performance. Finally, if \eqref{eqn:MCG optimization} was modified with terminal cost and constraint to ensure infinite horizon constraint enforcement, then the corresponding modified version of \eqref{eqn:NN-MCG optimization} would posses the same property as well. However, this extension is not the focus of this paper.
\end{remark}
\vspace{-0.2em}
A challenge with the approach described above is the computation of $M$. The analytical computation of $M$ can be quite tedious since it requires evaluation of the second-order sensitivity function of $y$ with respect to $v$, which can be difficult to obtain and may lead to overly conservative response.
A more practical and computationally efficient method to find $M$ is to numerically ``tune'' the NN-MCG by replacing \( M \) in \eqref{eqn:y upper bound replacment} with some \( \bar{M} \), that is:
\begin{flalign}
   \begin{split}
       & y_{tn}(j) + \sum_{k=0}^{j} S_y(j, k)  \left(v_t(k) - v_{tn}(k)\right)\\
       & + \frac{\bar{M}}{2} \sum_{k=0}^{j} \left(v_t(k) - v_{tn}(k)\right)^2\le \varepsilon
   \end{split}
\label{M_bar condition}
\end{flalign}
and calibrate $\bar{M}$ using data. We refer to this approach as $\bar{M}$-tuning. The following remark addresses the calibration of $\bar{M}$ in NN-MCG.
\begin{remark} 
In the $\bar{M}$-tuning method, $\bar{M}$ is determined iteratively through hardware experiments or simulations with NN-MCG in the loop. A reference signal of sufficiently long duration is selected to ensure $\bar{M}$ is tuned for the worst-case scenario. The method begins with \( \bar{M} = 0 \). At each timestep, \eqref{eqn:NN-MCG optimization} is solved, where $M$ is replaced by $\bar{M}$. If there is constraint violation, $\bar{M}$ is increased slightly and the process continues until no constraint violation is observed. Clearly, there is a trade-off between tracking performance and constraint satisfaction. As \( \bar{M} \) increases, tracking performance degrades while the  constraints become tighter. Conversely, reducing \( \bar{M} \) improves tracking performance but may result in constraint violations that are no longer considered as ``soft''. If tuning \( \bar{M} \) does not yield a satisfactory balance between tracking and constraint violation, then additional data collection and training of a more generalized NN is recommended.

\label{re:M_tuning}
\end{remark}

\begin{remark} For a single-input multi-output system (\( y(t) \in \mathbb{R}^{n_y} \)), Eq. \eqref{eqn:y upper bound replacment} in NN-MCG can be reformulated into \( n_y \) individual constraints, one for each output. Each constraint is based on the nominal prediction trajectory, the sensitivity function of the output, and a calibration parameter specific to that output.
\label{re:multiple outputs}
\end{remark}

\section{NN-MCG for constraint management of the Fuel Cell (FC) system}
\label{sec:Implementation of the modified NNRG on the FC air-path system}

 FCs are energy conversion devices that generate electricity via electrochemical reactions, typically between hydrogen and oxygen.  
In this section, the proposed NN-MCG is utilized for  constraint management of the FC air-path subsystem. The plant model used in this study is a reduced third-order  nonlinear, spatially averaged,  model of a FC stack with its auxiliaries \cite{talj2009experimental}. In this model, the fuel cell stack consists of $381$ cells and can supply up to $300$ A of current. This low-order model is derived from a nine-state full-order system developed in \cite{pukrushpan2004control}, and is suitable for studying the air-path side of the FC system. Below, we first introduce the closed-loop FC airpath system, i.e., system \eqref{eqn:NL_dynamic}, and show that without a governor scheme, system constraints are violated, which motivates the need for the NN-MCG. 

\subsection{Air-path system model of FC}
\label{sec:Air-path system model for FC}
The dynamics of the FC air-path system can be described by three dynamic states $x={\left[p_{\text{ca}},{\omega }_{\text{cp}},p_{\text{sm}}\right]}^\top$, consisting of the cathode air pressure, the angular velocity of the compressor motor, and the pressure of air in the supply manifold, respectively. The state-space representation of a reduced third-order FC model is presented in \cite{talj2009experimental}. 

Control of the air-path system aims to maximize net power while preventing oxygen starvation and avoiding compressor surge and choke  across a range of load conditions \cite{vahidi2006constraint}. The details of the feedforward and feedback controller design for the air-path system are given in \cite{pukrushpan2004control}. 

The constrained outputs of the air-path subsystem are ${y={\left[{\lambda }_{O_2},W_{\text{cp}},\ p_{\text{sm}}\right]}^\top}$, which  define the surge, choke, and oxygen starvation constraints respectively as follows \cite{vahidi2006constraint}:
\begin{flalign}
    & \frac{p_{\text{sm}}}{p_{\text{atm}}} \le 50 W_{\text{cp}} - 0.1, \ 
    \frac{p_{\text{sm}}}{p_{\text{atm}}} \ge 15.27 W_{\text{cp}} + 0.6, \
    \lambda_{O_2} \ge 1.9 &
    \label{eqn:FC_constraints}
\end{flalign}
where
\begin{flalign*}
    \begin{split}
        \lambda _{O_{2} } &=\frac{W_{O_{2} ,\text{ca,in}} }{W_{O_{2} ,\text{ca,rct}} }\\ 
        W_{O_{2} \text{ca,in}} &=k_{\text{ca,in}} \frac{x_{O_{2} ,\text{atm}} }{1+\omega _{\text{atm}} } \left(p_{\text{sm}} -p_{\text{ca}} \right)\\ 
        W_{O_{2} ,\text{ca,rct}} &=\frac{n_{\text{cell}} M_{O_{2} } }{4F} I_{\text{st}}
    \end{split}
\end{flalign*}
\noindent in which $W_{O_2,\text{ca,in}}$ is the oxygen flow rate into the cathode, $W_{O_2,\text{ca,rct}}$ is the flow rate of the oxygen reacted in the cathode, $p_{\text{atm}}$ is the atmospheric pressure, and $k_{\text{ca,in}}$ is the constant for the cathode inlet nozzle, $I_{\text{st}}$ is the stack current, $W_{\text{cp}}$ denotes the compressor mass flow rate, $n_{\text{cell}}$ is the number of cells in the stack, $F$ is the Faraday number, $\lambda_{O_2}$ is the oxygen excess ratio (OER), ${\omega }_{\text{atm}}$ is the humidity ratio of the \mbox{atmospheric} air, $x_{O_2,\text{atm}}$ is the atmospheric oxygen mass fraction. For further details and numerical values of the physical parameters, please refer to \cite{pukrushpan2004control}.

\begin{figure} 
\centering
\includegraphics[width=0.45\textwidth]{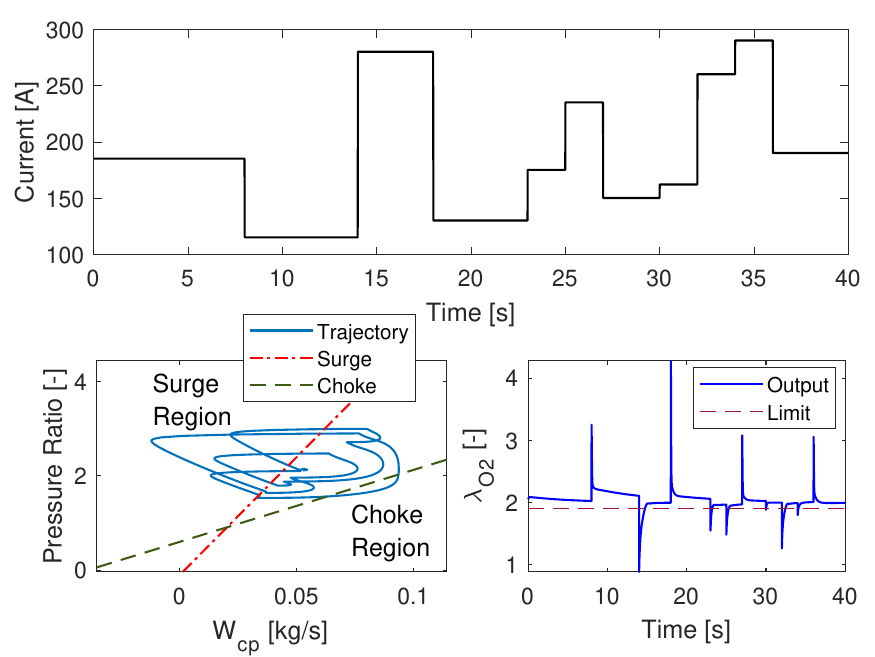}
\caption{Simulation response without constraint management: Current, OER, and compressor state trajectories for step changes in current demand. The dashed lines indicate system constraints. The surge region lies to the left of the dashed red line (surge boundary) and the choke region lies to the right of the dashed green line (the choke boundary). The region below $\lambda _{O_{2} } = 1.9$ on the OER plot indicates oxygen starvation and should be avoided for durable operation of  FC.}
\label{fig:FC_violation}
\vspace{-1.2em}
\end{figure}

Fig.~\ref{fig:FC_violation} illustrates the OER and the compressor trajectory of the closed-loop model (without any constraint management strategy) during a series of step-up and step-down changes in stack current $I_{\text{st}}$. 
The figure reveals that the surge constraint may be violated during current step-down, while the OER and choke constraints may be violated during current step-up. Since fast changes in current result in constraint violations, we employ a current governor (also known as a load governor), based on the NN-MCG strategy, to enforce the constraints on the OER output and compressor states.

\subsection{Data collection and training of the neural network}
\label{sec:NN for load control}
\begin{figure}
\centering
\includegraphics[width=0.45\textwidth]{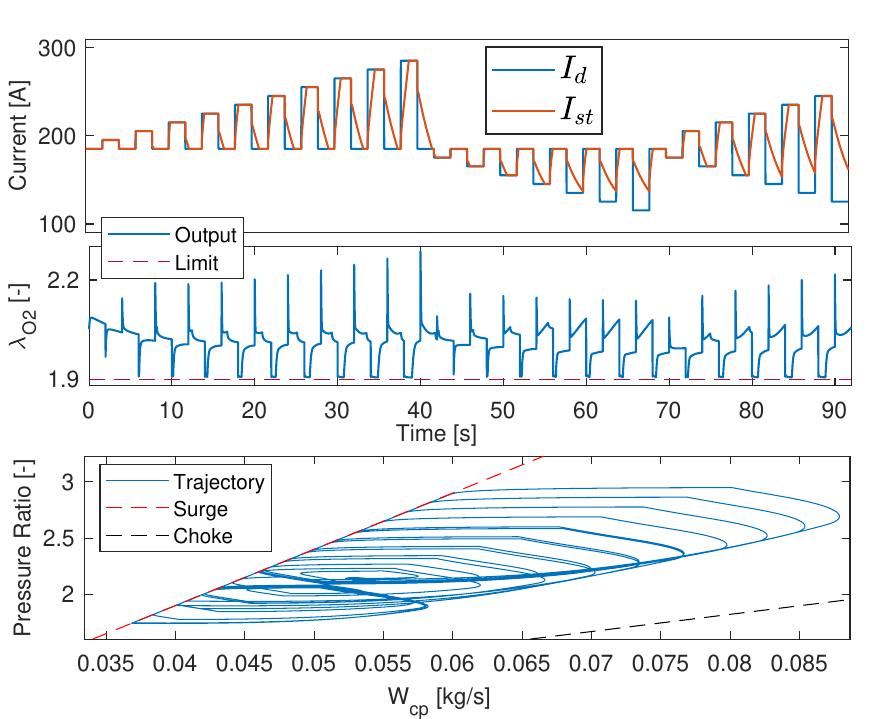}
\caption{Simulation results of the MCG applied to the FC system using a comprehensive reference profile for data collection.}
\label{fig:Data collection}
\vspace{-1.2em}
\end{figure}

Recall that the NN in the NN-MCG algorithm is trained to approximate the MCG control law through a training dataset. The training dataset is a collection of input and output data from the closed-loop system controlled by the MCG. Here, the inputs to the MCG (and therefore the NN) are the state $x(t)$ (of both the airpath and the controller) and the desired reference $r(t)=I_\text{d}(t)$. The output of the MCG is the modified current demand (stack current) over the prediction horizon, $V_t = [I_{\text{st}}(t|t), \dots, I_{\text{st}}(t+N|t)]^\top$. 
NNs achieve better generalization when trained on a dataset that covers a broad range of operating conditions. Thus, we simulate the closed-loop FC system with MCG in the loop using the reference profile shown in Fig.~\ref{fig:Data collection}. 

To incorporate the OER and surge constraints in \eqref{eqn:FC_constraints} into the MCG formulation, they are softened (similar to  \mbox{$y_t(j) \le \varepsilon$} in  \eqref{eqn:MCG optimization}). Specifically,
\begin{flalign}
    & 1.9 - y_{1t}(j) \leq \varepsilon_{1}, \quad
    \frac{y_{3t}(j)}{p_{\text{atm}}} - 50 y_{2t}(j) + 0.1 \leq \varepsilon_{2}
    \label{eqn:MCG constraint in FC}
\end{flalign}
where $y_1 = {\lambda }_{O_2}, y_2 = W_{\text{cp}}$, and $y_3 = p_{\text{sm}}$ as before. Note that \eqref{eqn:MCG constraint in FC} excludes the choke constraint, as for the given model, satisfying the OER constraint drives the compressor trajectory away from its choke boundary. The penalty weight  for both slack variables in \eqref{eqn:MCG constraint in FC} is chosen as \(\rho = 10^8\), and the penalty weight on the input rate of change is set to \(\rho_s = 10^4\). Here, the NLP optimization problem in \eqref{eqn:MCG optimization} is solved using {\ttfamily fmincon} in MATLAB, which leverages the Interior Point algorithm. For this simulation, the sampling frequency is set to $100$ Hz, and the output prediction horizon is set to $0.1$ s, corresponding to $N = 11$ and resulting in $9200$ data points.

To approximate the MCG control law using the collected dataset, we exploited a feedforward architecture of NN, configured with a single hidden layer that employs the hyperbolic tangent sigmoid function as the hidden layer activation function and a linear function in the output layer. Training of the NN is performed using the collected data, utilizing the command \ttfamily{fitnet} \normalfont of MATLAB. The cost function, the training algorithm, and the stopping criteria of the algorithm are set to the recommended default values of {\ttfamily{fitnet}}. After multiple trials with different parameters, we selected the best-performing NN comprising 10 neurons in the hidden layer to approximate the MCG control law. The RMSE of the trained NN is 0.656 [A], evaluated on the dataset used during the training process, indicating successful training of the NN. It is notable that as few as 10 neurons were sufficient to yield an accurate approximation of the MCG. 

The next step in the design of NN-MCG is computing the sensitivity functions $S_y(j,k)$. The next subsection presents an analysis of these functions for the FC system.

\subsection{Analysis of the sensitivity functions}

\begin{figure}
\centering
\includegraphics[width=0.5\textwidth]{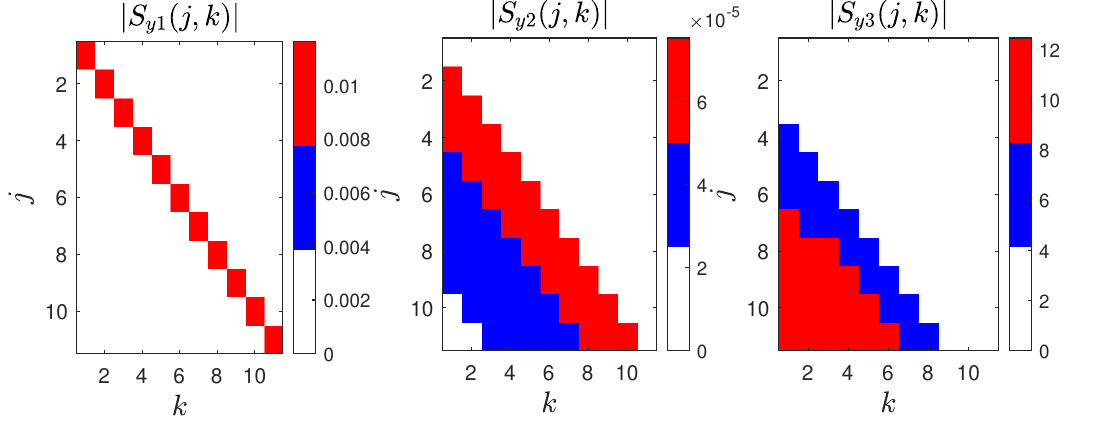}
\caption{Sensitivity functions of the constrained outputs at a representative operating point.}
\label{fig:Sensitivity plot}
\vspace{-1.2em}
\end{figure}

Recall from Section~\ref{sec:sensitivity} that obtaining a truncated Taylor series expansion of the output prediction requires a nominal input sequence (obtained from NN inference), as well as the sensitivity of the constrained outputs with respect to inputs around the nominal values. 
The plots in Fig.~\ref{fig:Sensitivity plot} 
show the sensitivity functions $S_{y_i}(j,k)$ defined in \eqref{eqn:y_sensitivity} for the three outputs, $y_1$, $y_2$, and $y_3$. 
Oxygen excess ratio, \( y_1 \), exhibits the strongest sensitivity to the current value of the input \( v_t(j) \), as indicated by the dominant diagonal in \( |S_{y1}(j,k)| \). 
In contrast, the compressor flow rate, \( y_2 \), demonstrates significant sensitivity to inputs from the preceding two to three timesteps, \( v_t(j-1) \), \( v_t(j-2) \), and potentially \( v_t(j-3) \).
Notably, the supply manifold pressure, \( y_3 \), displays the most pronounced sensitivity to past inputs, with \( |S_{y3}(j,k)| \) indicating a strong influence from inputs extending several timesteps back. 
These observations align well with the physical behavior of the system, highlighting the importance of capturing the sensitivity to past as well as current inputs in Eq. \eqref{eqn:taylor expansion}. Note that ignoring past inputs (i.e., changing the summation lower bound from 0 to $j$ in \eqref{eqn:taylor expansion}) results in a poor Taylor expansion, large truncation errors, and degraded NN-MCG performance.
\vspace{-1.2em}
\subsection{NN-MCG for load control}
\label{sec:NN-MCG for load governor}
In this section, the NN-MCG algorithm is employed as a load governor for constraint enforcement in FC system. The NN-MCG uses the same trained NN from the previous subsection.
The FC system has multiple constrained outputs. Therfore, according to Remark~\ref{re:multiple outputs}, for the given multi-output system, we can rewrite the nonlinear constraints in \eqref{eqn:MCG constraint in FC} over the prediction horizon as the following quadratic constraints:
\begin{flalign}
    \begin{split}
        &\frac{M_{1}}{2} \sum_{k=0}^{j} \left(v_t(k) - v_{tn}(k)\right)^2 - \sum_{k=0}^{j} S_{y_1}(j, k) \left(v_t(k) - v_{tn}(k)\right)\\
        &\quad -y_{1tn}(j) +1.9 \leq \varepsilon_1\\
        &\frac{M_s}{2}\sum_{k=0}^{j}(v_t(k) - v_{tn}(k)^{2} + \sum_{k=0}^{j}\left(S_{y_{3}}(j,k) - 50p_{\text{atm}} S_{y_{2}}(j,k)\right) \\
        &\left(v_t(k) - v_{tn}(k)\right) + {y}_{3tn}(j) - 50p_{\text{atm}} {y}_{2tn}(j) + 0.1p_{\text{atm}} \leq \varepsilon_2
    \end{split}
\label{eqn:quadratic constraints of FC}
\end{flalign}
where $M_{s}=M_{3} + 50p_{\text{atm}}M_2$, and $M_{i} \ge 0$, $i=1,2,3$ is a constant for the \(i\)-th output of the FC system such that \( \left| \frac{d^2 y_i}{dv^2} \right| \leq M_i \) for all $x$ and $v$ within the operating range of the FC system. Please see \cite{ayubirad2024machine} for details on computing the nominal state equations and the sensitivities of the FC model. As discussed in Section~\ref{sec:sensitivity}, 
for practical implementation of the NN-MCG, $M_{1}$ and $M_{s}$ in  \eqref{eqn:quadratic constraints of FC} are respectively substituted with the parameters $\bar{M}_{\lambda_{O_{2}}}$ and $\bar{M}_{s}$. 
Calibration of these parameters is performed using the reference profile from Fig.~\ref{fig:FC_violation}, where closed-loop simulations with NN-MCG are conducted, and the parameters are tuned to achieve a satisfactory balance between tracking and constraint violation (see Remark~\ref{re:M_tuning}). The QCQP optimization problem of NN-MCG is solved using the Gurobi solver in MATLAB, with the slack penalty weights in \eqref{eqn:quadratic constraints of FC} set to \( \rho_1 = 10^8 \) and  ${\rho_2 = 10^3}$, and the penalty weight on the input rate of change set to  ${\rho_s = 10^5}$ .

\subsection{Drive-cycle simulation and computational analysis}
\label{Drive-cycle simulation}
\begin{figure}
    \centering
   \begin{subfigure}{0.4\textwidth}
        \centering
        \includegraphics[width=\textwidth]{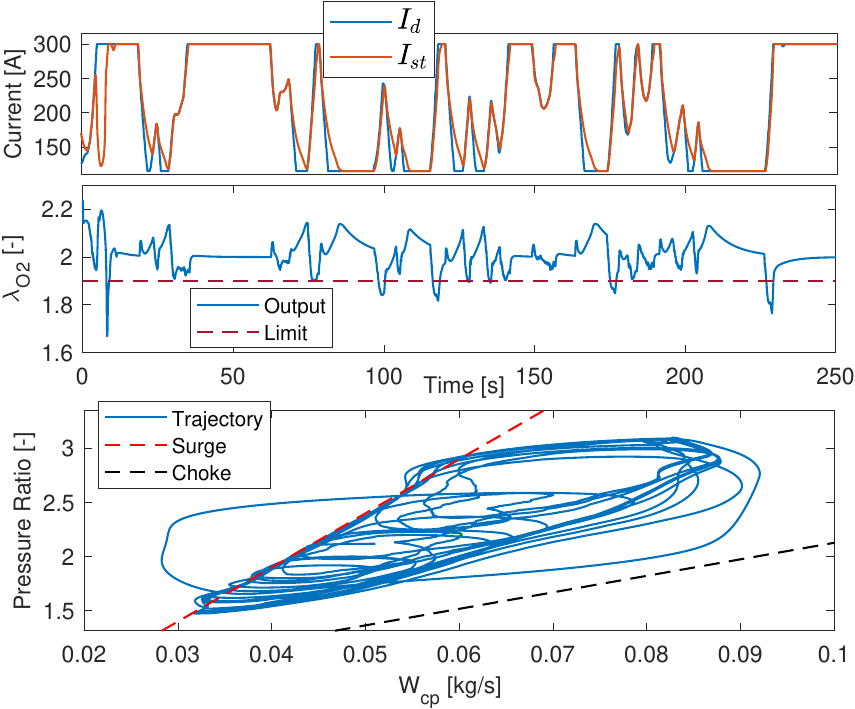}
        \caption{Naive NN}
    \end{subfigure}
    \begin{subfigure}{0.4\textwidth}
        \centering
        \includegraphics[width=\textwidth]{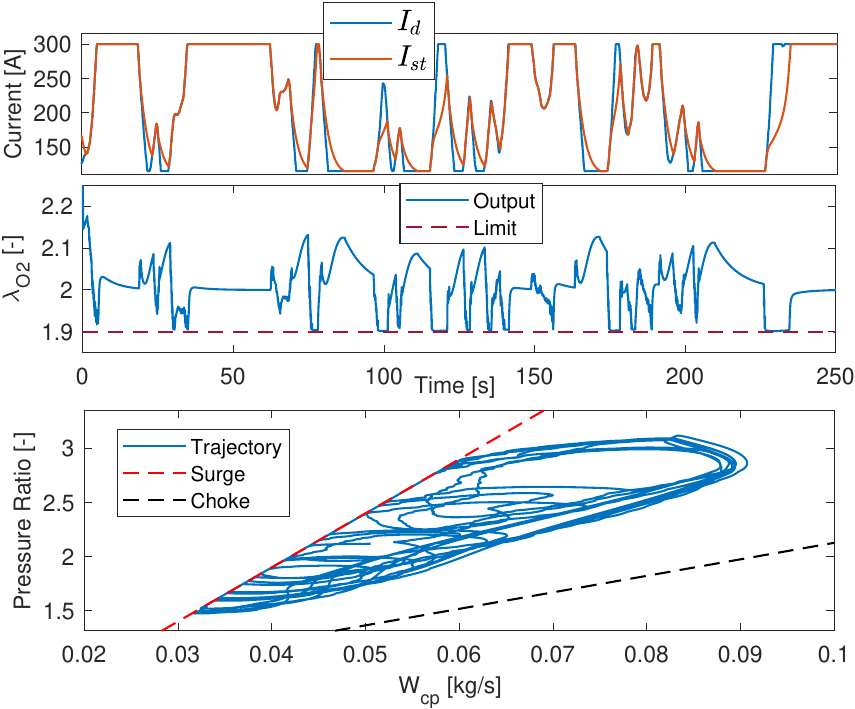}
        \caption{NN-MCG}
    \end{subfigure} 
    \caption{Naive NN and NN-MCG on Dynamic test}
    \label{fig:drive cycle}
    \vspace{-1.0em}
\end{figure}

In this section, we evaluate the performance of the NN-MCG with $\bar{M}$-tuning on a realistic drive-cycle. The current demand used for this validation is representative of the FC system power needed for the completion of a standard drive-cycle, referred to as “Dynamic Test”, as shown Fig.~\ref{fig:drive cycle}. 
To evaluate the effectiveness of our approach, the NN-MCG algorithm is applied to the drive cycle simulation and compared with the naive NN (see Remark \ref{re:NN_fail}). We also simulated the FC system on the same drive cycle using MCG. Due to space limitations,  simulation results of MCG are not shown. The key findings from comparing the naive NN results in Fig. ~\ref{fig:drive cycle}(a), the NN-MCG results in Fig. ~\ref{fig:drive cycle}(b), and the MCG results are: 
\begin{itemize}
    \item The naive NN, even trained on such a comprehensive dataset, fails to produce a load response that satisfies the constraints at all times due to inherent approximation error of the NN. 
    \item NN-MCG effectively produces a load response that achieve a satisfactory balance between tracking performance and constraint enforcement and aligns well with the load response of the MCG, achieving an RMSE of $0.985$ [A], with maximum slack values of ${\varepsilon_1=3.63 \times 10^{-7}}$ and $\varepsilon_2 = 1.54 \times 10^{-4}$.
\end{itemize}

Additionally, we present a computational comparison between the  MCG, naive NN, and the proposed NN-MCG. All simulations are conducted in MATLAB on a laptop equipped with a 12\textsuperscript{th} Gen Intel(R) Core(TM) i9-12900H @ 2.90 GHz processor and 16 GB of memory. 
Each method is simulated five times to reduce variability from background processes. For each timestep, the minimum computational time across the runs is recorded. The average execution time is computed as the sum of these minimum times divided by the total timesteps, while the worst-case execution time is the maximum of these minimums. Both metrics are reported in Table~\ref{tab:Performance of different load governors}.
\begin{table}
\centering
\begin{threeparttable}
\caption{\\ \uppercase{Performance of Different Load Governors}}
\footnotesize 
\begin{tabular}{|c|c|c|c|}
\hline
\multirow{2}{*}{\textbf{Case}} & \multicolumn{1}{c|}{Average} & \multicolumn{1}{c|}{Worst-case} & \multicolumn{1}{c|}{Satisfied} \\ 
& \multicolumn{1}{c|}{execution time [ms]} & \multicolumn{1}{c|}{ execution time [ms]} & \multicolumn{1}{c|}{performance?} \\ \hline
Naive NN & $2$ & $3.4$ & No \\ \hline
MCG & $32.9$ & $146$ & Yes \\ \hline
NN-MCG & $4.5$ & $6.3$ & Yes \\ \hline
\end{tabular}
\label{tab:Performance of different load governors}
\end{threeparttable}
\end{table}
According to the results in Table~\ref{tab:Performance of different load governors}, the following points about each method can be concluded:
\begin{itemize}
    \item The MCG has the highest average and worst-case execution times, exceeding the sampling time and posing real-time execution challenges. 
    \item The naive NN has the lowest average and worst-case execution times (the worst case being within the sampling time).
    \item The NN-MCG framework strikes a balance with a moderate average execution time of $4.5$ ms, a worst-case execution time of $6.3$ ms, and a performance close to the MCG. 
\end{itemize}

The high computation time of MCG stems from online NLP optimization, which requires multiple iterations, each involving at least one simulation of the system. Meanwhile, the naive NN is trained offline, with its online computation limited to NN inference. The NN-MCG strikes a balance between full online computation and purely offline naive NN and exhibits a worse-case execution time roughly 23 times smaller than the MCG. The additional computational cost of NN-MCG over naive NN is due to a single system simulation per timestep and solving a QP/QCQP. However, this cost remains significantly lower than the online NLP optimization in MCG.

\section{Conclusions}
In this paper, we presented a novel neural network-based multi-timestep command governor (NN-MCG) for constraint management of nonlinear systems. The proposed approach addresses the computational complexity of traditional multi-timestep command governor (MCG) by using an offline trained neural network (NN) to approximate the optimal input sequence of MCG, followed by a modification using an online sensitivity-based strategy to further enhance the constraint enforcement capability of the NN. The effectiveness of the proposed NN-MCG was demonstrated through its application as a load governor for an automotive fuel cell system. Results showed that the online computational footprint of NN-MCG is significantly smaller than that of MCG, while achieving almost the same performance, making it more suitable for real-time implementation on fast nonlinear systems. 
Future work will explore the integration of advanced machine learning models in NN-MCG and evaluate the proposed method on other dynamical systems.
\label{sec:conclusions}

\bibliographystyle{IEEEtran}
\bibliography{conference_101719}

\end{document}